\documentclass[a4paper,11pt]{article}
\usepackage{pos}

\usepackage{gensymb}
\usepackage{natbib}
\title{\boldmath EUSO-SPB2 Fluorescence Telescope Calibration and Field Tests }
 \ShortTitle{EUSO-SPB2 Fluorescence Telescope Calibration and Field Tests }

\author*[a]{Viktoria Kungel}
\author[b]{Matteo Battisti}
\author[a]{George Filippatos}
\author[a]{Tobias Heibges} 
\author[c]{Evgeny Kuznetsov}
\author[d]{Marco Mese}
\author[e]{Stephan S. Meyer}
\author[f]{Etienne Parizot}
\author[d]{Valentina Scotti}
\author[a]{Partrick Sternberg}
\author[a]{Lawrence Wiencke}

\affiliation[a]{Colorado School of Mines,\\
1500 Illinois St, Golden, USA}

\affiliation[b]{Universitá di Torino,\\
V. P. Giuria 1, Turin, Italy}

\affiliation[c]{University of Alabama in Huntsville,\\
301 Sparkman Dr NW, Huntsville, USA}

\affiliation[d]{Università degli Studi di Napoli Federico II,\\
Corso Umberto I 40, Naples, Italy }

\affiliation[e]{University of Chicago,\\
5801 S. Ellis Ave., Chicago, USA}

\affiliation[f]{APC, Univ Paris Diderot, CNRS/IN2P3, CEA/Irfu,\\
13 rue Watt, Paris, France}

\onbehalf{for the JEM-EUSO Collaboration\\[-1mm]{\normalsize \normalfont (a complete list of authors can be found at the end of the proceedings)}}
\emailAdd{kungel@mines.edu}
\abstract{The Extreme Universe Space Observatory on a Super Pressure Balloon 2 (EUSO-SPB2), successfully launched from Wanaka, New Zealand on May 13, 2022, is a precursor for a space-based astroparticle observatory such as the Probe Of Extreme Multi-Messenger Astrophysics (POEMMA). EUSO-SPB2 flew two custom telescopes. Both have UV/UV-visible sensitivity and feature Schmidt optics. The Fluorescence Telescope (FT) measures ultra-high energy cosmic rays by looking down. The \v{C}erenkov Telescope (CT) searches for neutrino signatures by looking toward Earth’s limb. The two telescopes each have a 1 m diameter entrance pupil and segmented glass mirrors that collect light from extensive air showers at the PeV and EeV-scale. \\
Here we describe the FT telescope optics together with the results of the FT field tests at the Utah Telescope Array (TA) site from August/September 2022. The FT recorded the night sky background, lasers, and artificial point sources. The field tests included an absolute photometric calibration of the FT telescope that is compared to a piece-wise laboratory calibration.
}

\ConferenceLogo{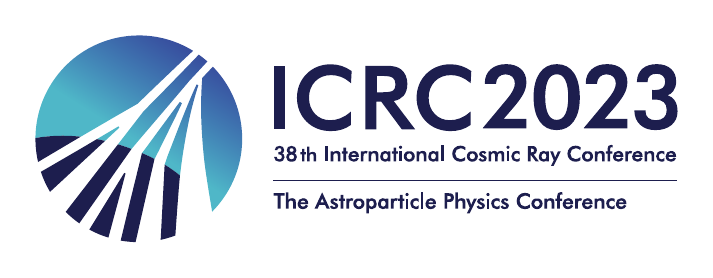}

\FullConference{%
38th International Cosmic Ray Conference (ICRC2023)\\
  26 July - 3 August, 2023\\
  Nagoya, Japan}

\begin{document}
\maketitle

%\section{...}
%Some text with ref. \cite{minieuso}

\section{EUSO-SPB2 Mission}
\begin{figure}[ht]
\centering
\includegraphics[width=1.\textwidth]{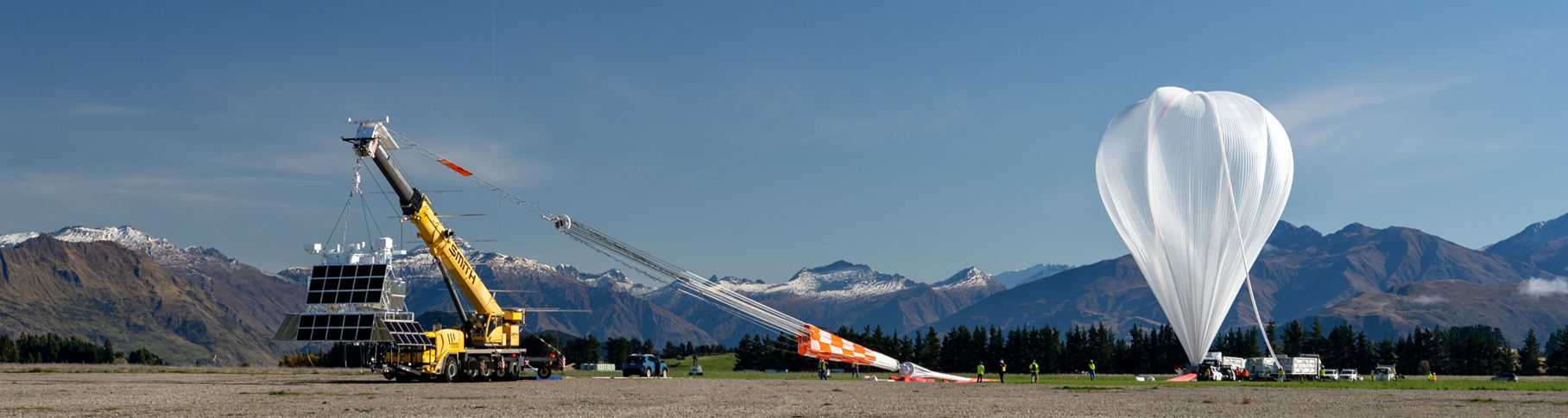}
\caption{Launch of EUSO-SPB2 from Wanaka, New Zealand, on May 13, 2022}\label{fig:SPB2}
\end{figure}

The Extreme Universe Space Observatory on a Super Pressure Balloon 2 (EUSO-SPB2), is a scientific and technological pathfinder designed to observe Ultra-High-Energy Cosmic Rays (UHECR), Very-High-Energy (VHE) cosmogenic neutrinos, and other astrophysical phenomena from the stratosphere, above most the atmospheric interference, aerosol, and light pollution. It is a follow-up mission to EUSO-SPB1 \cite{SPB1}, which was successfully launched in 2017, and a precursor for space-based astroparticle observatories such as the Probe Of Extreme Multi-Messenger Astrophysics (POEMMA) \cite{Olinto_2021}.
EUSO-SPB2 has two optical Schmidt-telescopes with a wide field of view, high-speed cameras with UV/UV-visible sensitivity, and an IR camera for cloud monitoring. \\
%EUSO-SPB2 launched successfully on May 13, 2022, from Wanaka (NZ) and flew a total of 5625 lbs at $\sim 100000$ feet over two nights with the moon down and collected a total of 57GB data, which are to be analyzed and looked for cosmic rays. 
EUSO-SPB2 successfully launched from Wanaka, New Zealand, on May 13, 2022, see Fig. \ref{fig:SPB2} and successfully reached a float altitude of 32 km.
The EUSO-SPB2 instruments turned on successfully and worked with 57 GB of data downloaded. Unfortunately, the balloon developed a bad leak and the flight terminated in the Pacific Ocean on May 14 after two nights of operations.
%Soaring at approximately 100,000 feet for two moonless nights, it carried a payload weighing 5625 lbs and collected 57 GB of data. The acquired data remains pending analysis.
Despite the short flight opportunity, it represents a significant milestone toward a space-based UHECR and VHE neutrino observatory.
%The successful launch and data collection represent significant milestones, advancing the research with near-space astroparticle detectors.

\section{EUSO-SPB2 Fluorescence Telescope (FT)}

The Fluorescence Telescope's (FT), Fig. \ref{fig:FT}, objective is to observe UHECR by looking down in the nadir from near orbital altitude for the first time and searching for upward-going event candidates.
The FT features a modified Schmidt design using a segmented spherical mirror with six mirror segments made of borosilicate glass.
The camera consists of three Photodetector Modules (PDM) with Multi-Anode Photo-Multiplier tubes (MAPMTs)\cite{MAPMT} and 6912 pixels. Additional optics include field-correcting or flattening lenses. All three PDMs have BG3 filters with a 300-420 nm UV light sensitivity. The following table gives an overview of the FT specifications:

\begin{figure}[ht]
\centering
\includegraphics[width=.4\textwidth]{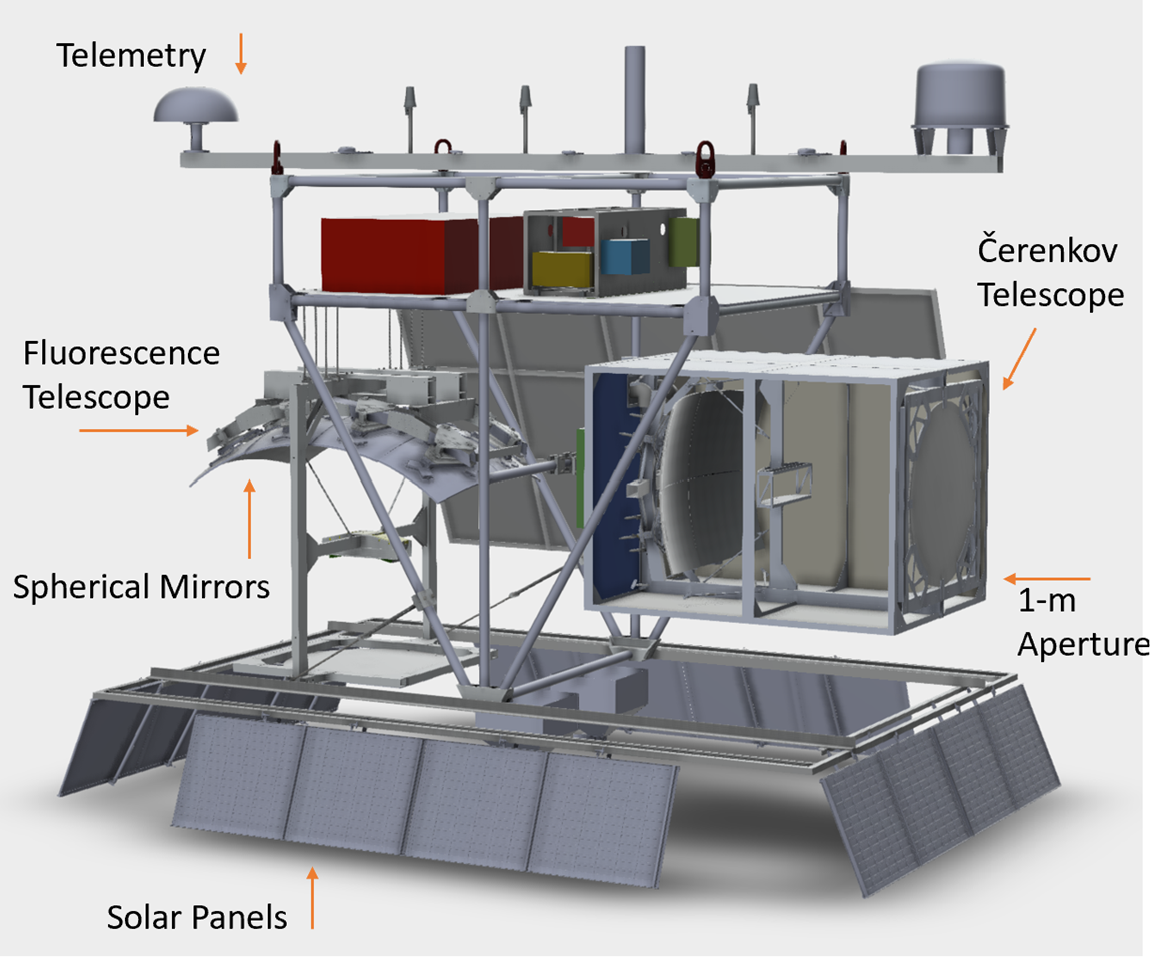}
\includegraphics[width=.59\textwidth]{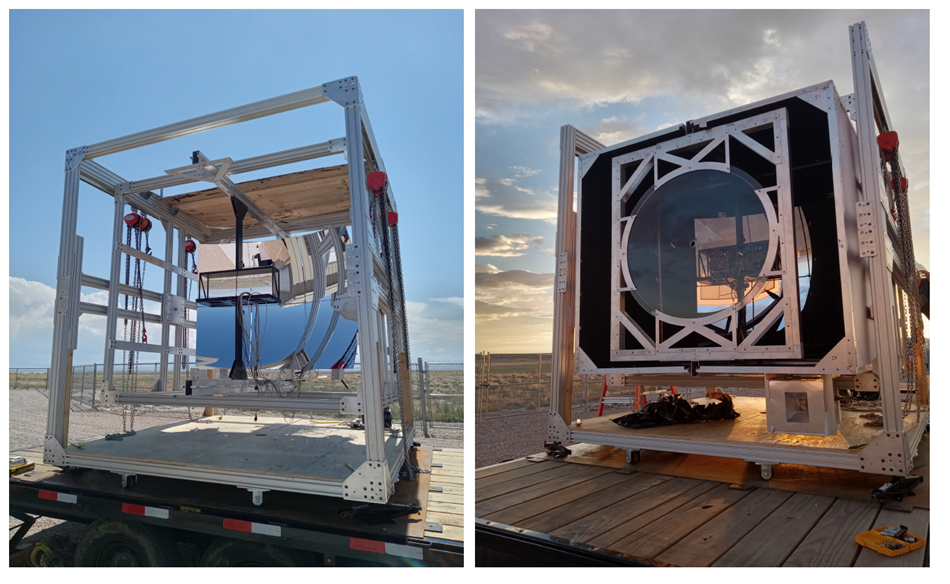}
\caption{The EUSO-SPB2 FT telescope model, half and full assembly. }\label{fig:FT}
\end{figure}

\begin{center}
\begin{tabular}{ l l }
\hline
Specification & FT  \\ 
 \hline
 Wavelength Sensitivity & UV 300-420 nm  \\  
 Energy threshold & EeV  \\  
 Sensor Type & MAPMT (Hamamatsu[]) \\
 FoV & $3\times (12.02\times 12.02 \pm 0.1) \degree$ \\
 %Pixel FoV & $(0.25\times 0.25) \degree$ \\
 Diameter ACP/ Image & 1 m //  0.79 m\\
 PSF/ Pixel Size & 1.75 mm // $3\time 3$ mm$^2$ \\
 \hline
\end{tabular}
\end{center}

\section{Characterization of the FT Optics}
\begin{figure}[ht]
\centering
\includegraphics[width=.47\textwidth]{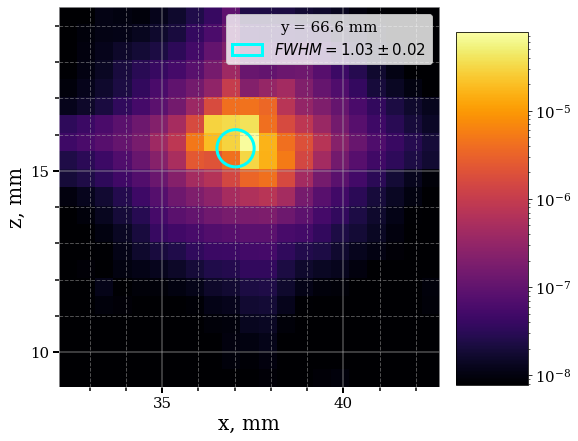}
\centering
\includegraphics[width=.5\textwidth]{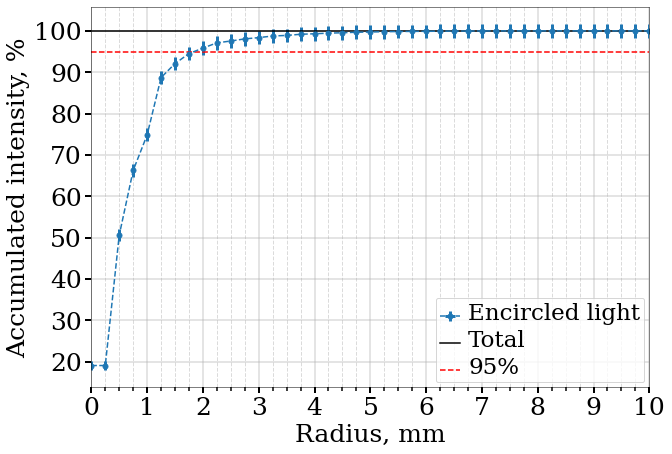}
\caption{PSF measurement of the FT in the lab with all six segments aligned to a common spot. A remote 3D-gantry system raster scans $1\times 1$ cm planes in 0.5 mm step size along the optical axis. The PSF at focus is 1.75 mm, where 95\% of the maximum intensity is encircled. }\label{fig:PSF}
\end{figure}

The characteristics of the FT Optics are the Point Spread Function (PSF) and the optical efficiency. Other than ground-based experiments, airborne experiments can not be adjusted or calibrated after launch. Thus, the PSF and efficiency have been measured prior to it in the lab in a dark room of less than a few nW/cm$^2$, which is significantly darker than the night sky. \\
For the lab measurements of the optics a customized 1-m parallel test beam has been fabricated to illuminate the entrance pupil of the telescope using a parabolic UV enhanced coated F3.5 mirror. The beam is parallel to a spread of $\sim 0.02 \degree$ which is small to a pixel FoV. \\
The PSF was measured with a 3D-gantry system that can be operated remotely outside the darkroom and has a photodiode attached that raster scans $1\times 1$ cm planes in 0.5 mm step size along the optical axis. Measuring the light at the focal plane after aligning all six mirror segments to a common spot gives a PSF of 1.75 mm, where 95\% of the maximum intensity is encircled, see Fig. \ref{fig:PSF}. This is within an FT pixel of $3\times 3$ mm and overlapping gaps.\\

\begin{figure}[ht]
\centering
\includegraphics[width=1.\textwidth]{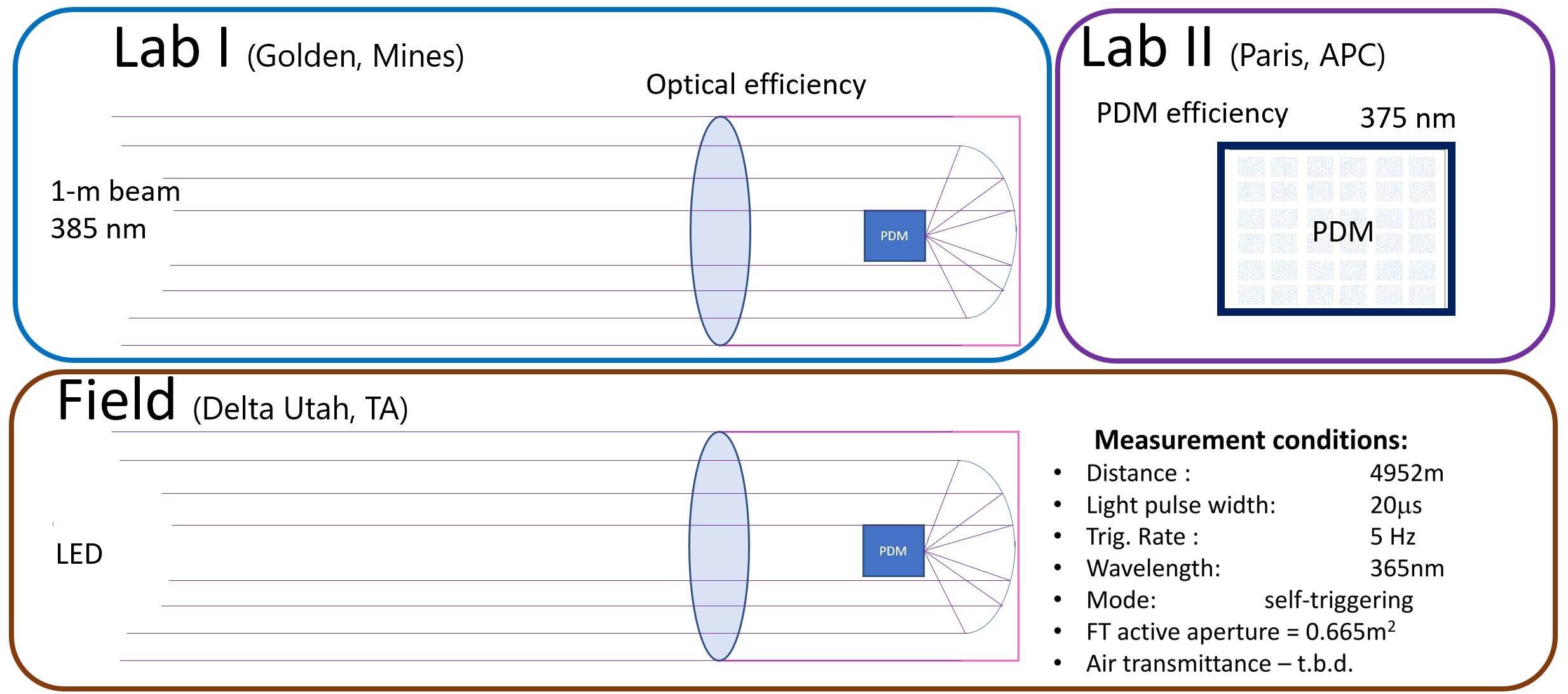}
\caption{An absolute photometric calibration in the lab compared to an end-to-end measurement in the field. The
optical efficiency folded with the PDM efficiency of the FT is $16 \pm 2$\%. The LED measurement gives an FT efficiency of 14.7\% in the far field and 20.7\% for PDM1/PDM2 or 23.9\% for PDM3 in the near field, without correction for aerosol attenuation. The EUSO-SPB2 efficiency is above EUSO-SPB1's.}\label{fig:cali}
\end{figure}

Comparing the light at the focus and the incoming light at the aperture of the FT gives the optical efficiency of the FT and is $45 \pm 1$ \%, which includes scattering and absorption at the Achromatic PMMA Corrector Plate (ACP), mirrors, BG3 filter, field flattener, and obscurations from the camera shelf. The optical efficiency combined with the PDM efficiency, which is $35\pm 4$\% leads to the end-to-end measurement of the whole telescope, Fig. \ref{fig:cali}, and the absolute calibration of the FT with a $16\pm 2$ \% efficiency. The EUSO-SPB2 efficiency is above the EUSO-SPB1 measurements of $10\pm 1$\% and is in agreement with the improved measurements of the trigger threshold and focusing in the field. \\

\section{FT Field Tests}
\begin{figure}[ht]
\centering
\includegraphics[width=1.\textwidth]{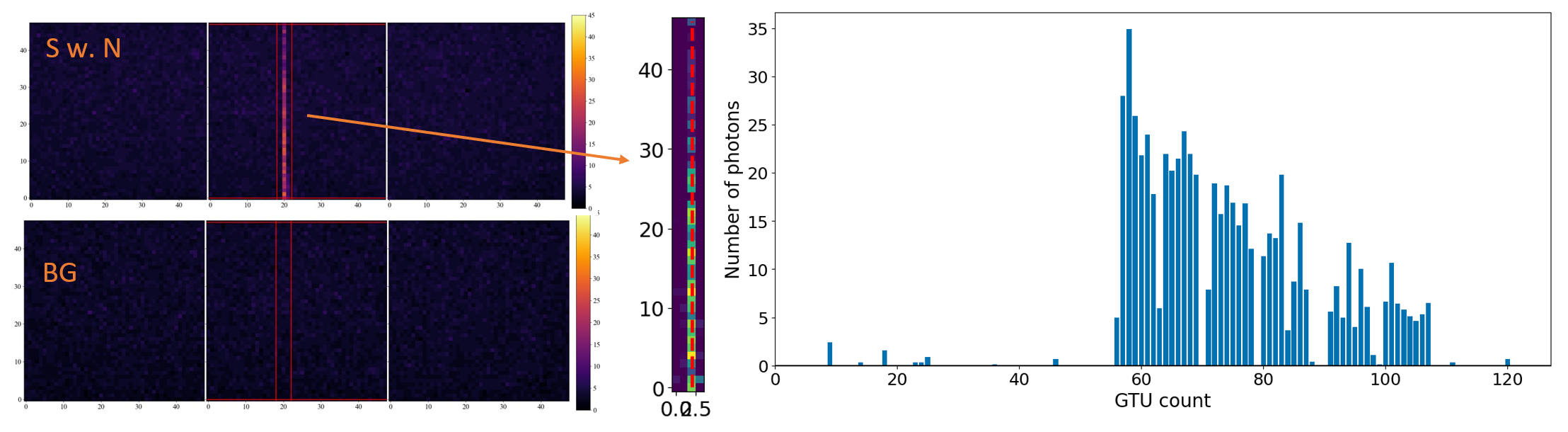}
\caption{Time integrated laser track at 24 km distance to the FT as seen in the FT camera. A track-finding algorithm is used to find laser-track events in the FT FoV and extract signals from noise.}\label{fig:track}
\end{figure}

A field campaign was carried out between August 19, 2022, and September 3, 2022, where the FT was driven out to a dark site in the Utah desert at the Telescope Array. 
A high-power UV laser system serves as a calibrated light source that simulates the signature of an Extensive Air Shower (EAS) of a primary particle with $10^{18}$ to $10^{21}$ eV. 
%The mobile laser system is enclosed in a utility trailer and can be uncoupled from the trailer when operating. 
The laser is a pulsed 355 nm frequency tripled YAG laser with an energy range of 200 $\mu$J to 90 mJ.
The beam-steering periscope can be aimed at any direction above the horizon with a pointing accuracy higher than $\pm 0.2$ \degree.
Remote operation of the laser enables its control from a distance, while each pulse is characterized using a monitoring probe.
A laser track-finding algorithm was written to identify laser events in the field data, see Fig. \ref{fig:track}, in addition to the time matching of triggered events in the FoV of the telescope and the auto-log files of the laser. \\

\begin{figure}[ht]
\centering
\includegraphics[width=1.\textwidth]{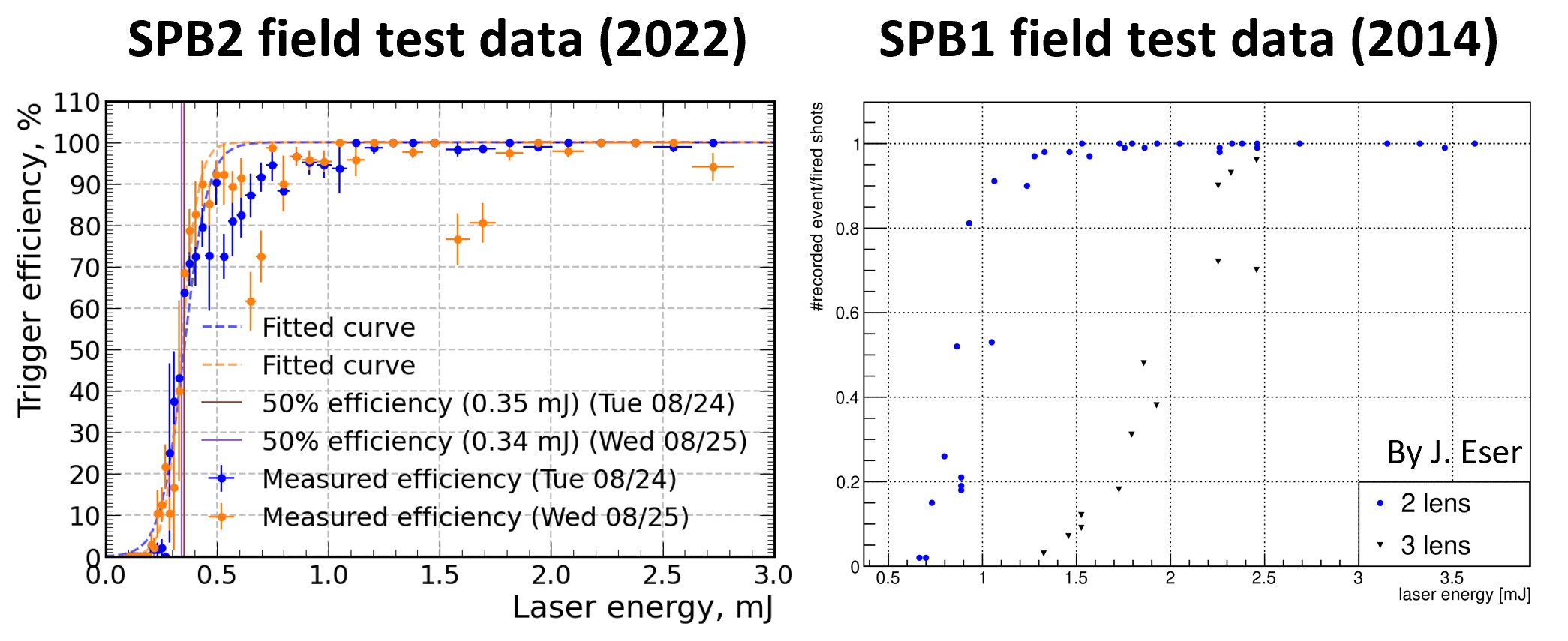}
\caption{EUSO-SPB2 trigger efficiency compared to EUSO-SPB1 with the same configuration in the field.}\label{fig:trigger}
\end{figure}

\newpage
One measurement maintains a consistent laser geometry while varying the laser energies, allowing for the determination of the trigger efficiency of the FT. 
The laser is positioned at a distance of 24 km from the FT, with an elevation angle of $45 \degree$ away from the telescope. The telescope itself is set at a $7 \degree$ elevation angle. Compared to EUSO-SPB1, 50\% of all laser shots were still seen at $(0.357\pm 0.006)$ mJ total laser energy with the FT, which is about the equivalence of a 9 EeV UHECR, while EUSO-SPB1 has no longer triggered on any laser tracks at that energy and had a trigger efficiency of $(0.94 \pm 0.02)$ mJ for a two lens configuration, see Fig. \ref{fig:trigger}. This is in agreement with the improved measurements of the efficiency and focus of EUSO-SPB2.

\begin{figure}[ht]
\centering
\includegraphics[width=1.\textwidth]{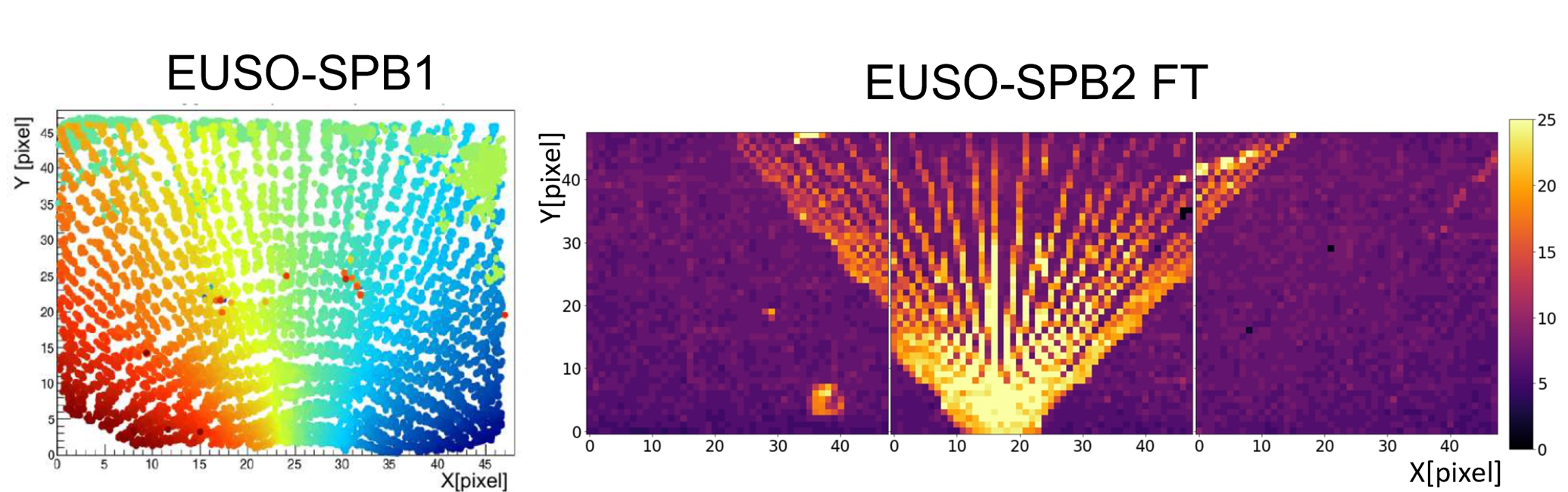}
\caption{Time-integrated laser pulses in the FoV of EUSO-SPB2 compared to EUSO-SPB1 with a distance of 24 km between laser and telescope. EUSO-SPB2 has three PDMs with a much better focusing.}\label{fig:sweep}
\end{figure}

In a second measurement, laser pulses with constant laser energy at 1.8 mJ are fired at 2 Hz with varying laser pointing angles in azimuth, enabling the evaluation of the FT's field of view (FOV). Both the telescope and laser remained at 24 km distance and with an elevation angle of $45 \degree$ for the laser and $7 \degree$ for the telescope. The EUSO-SPB2 FT laser sweep in 2022 compared to the same measurement with EUSO-SPB1 in the year 2017 can be seen in Fig. \ref{fig:sweep}. EUSO-SPB1 has only one camera compared to the 3 PDMs of the FT. It has a smaller FoV with $FoV_{SPB1} = (11.1 \pm 0.1)\degree$ and a pixel FoV with $0.23\degree$, compared to the EUSO-SPB2 FT with $FoV_{SPB2} = (12.02 \pm 0.1)\degree$ and a pixel FoV with $0.25\degree$.

\subsection{Other Measurements}

\begin{figure}[ht]
\centering
\includegraphics[width=1.\textwidth]{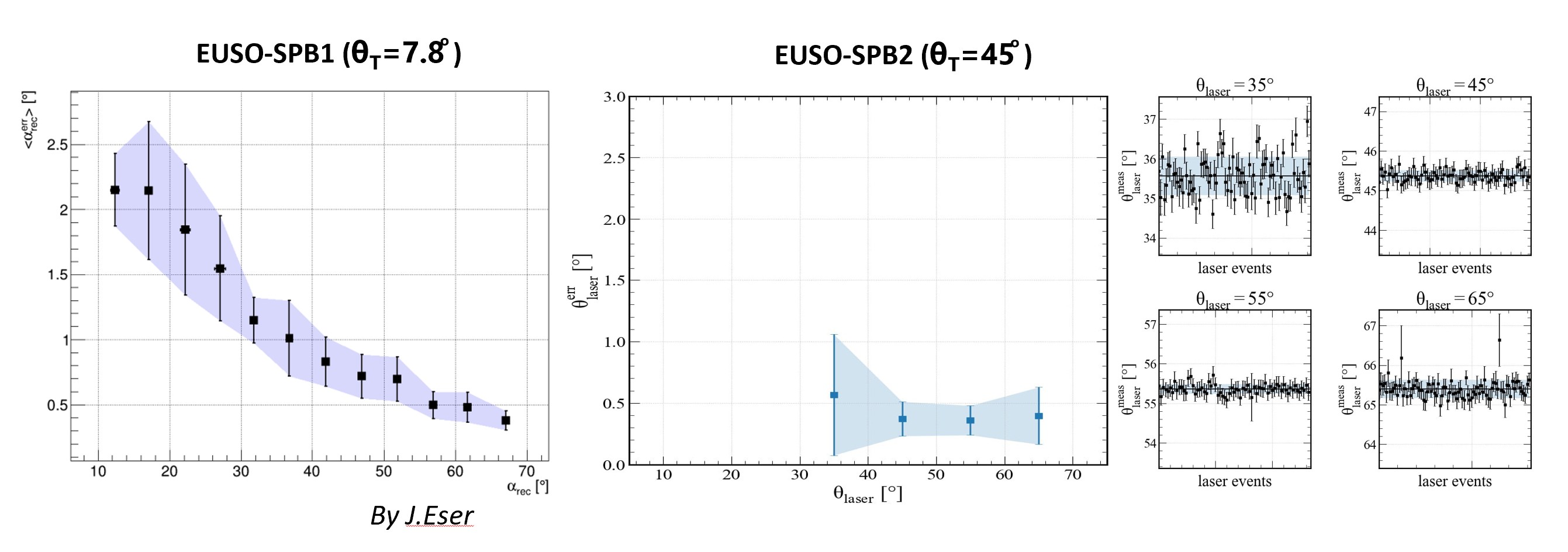}
\caption{The angular resolution measurements. EUSO-SPB1 has an elevation angle of 7\degree and sees a longer laser track in the FoV, while the FT of EUSO-SPB2 looks relatively high at 45\degree. }\label{fig:angle}
\end{figure}

\begin{figure}[ht]
\centering
\includegraphics[width=.45\textwidth]{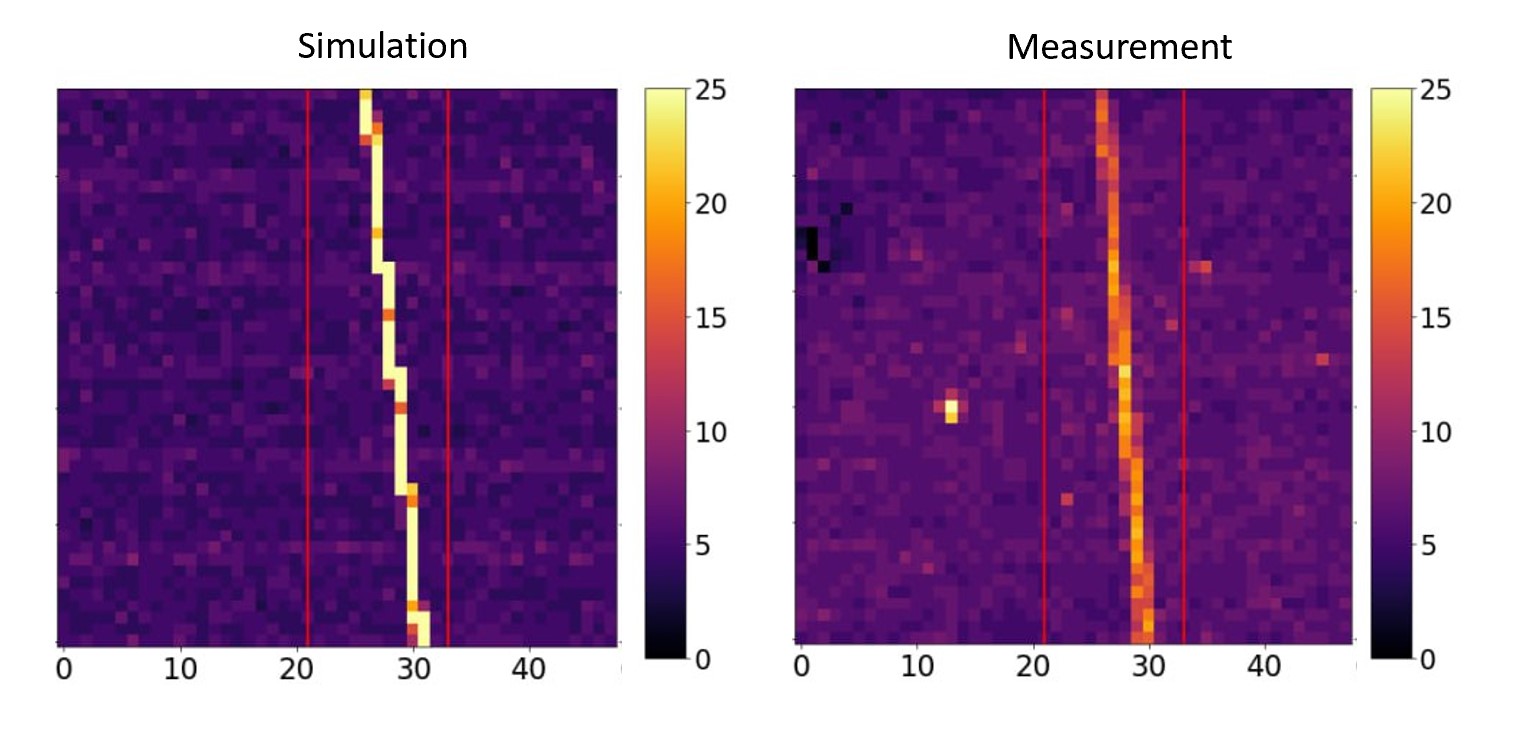}
\includegraphics[width=.4\textwidth]{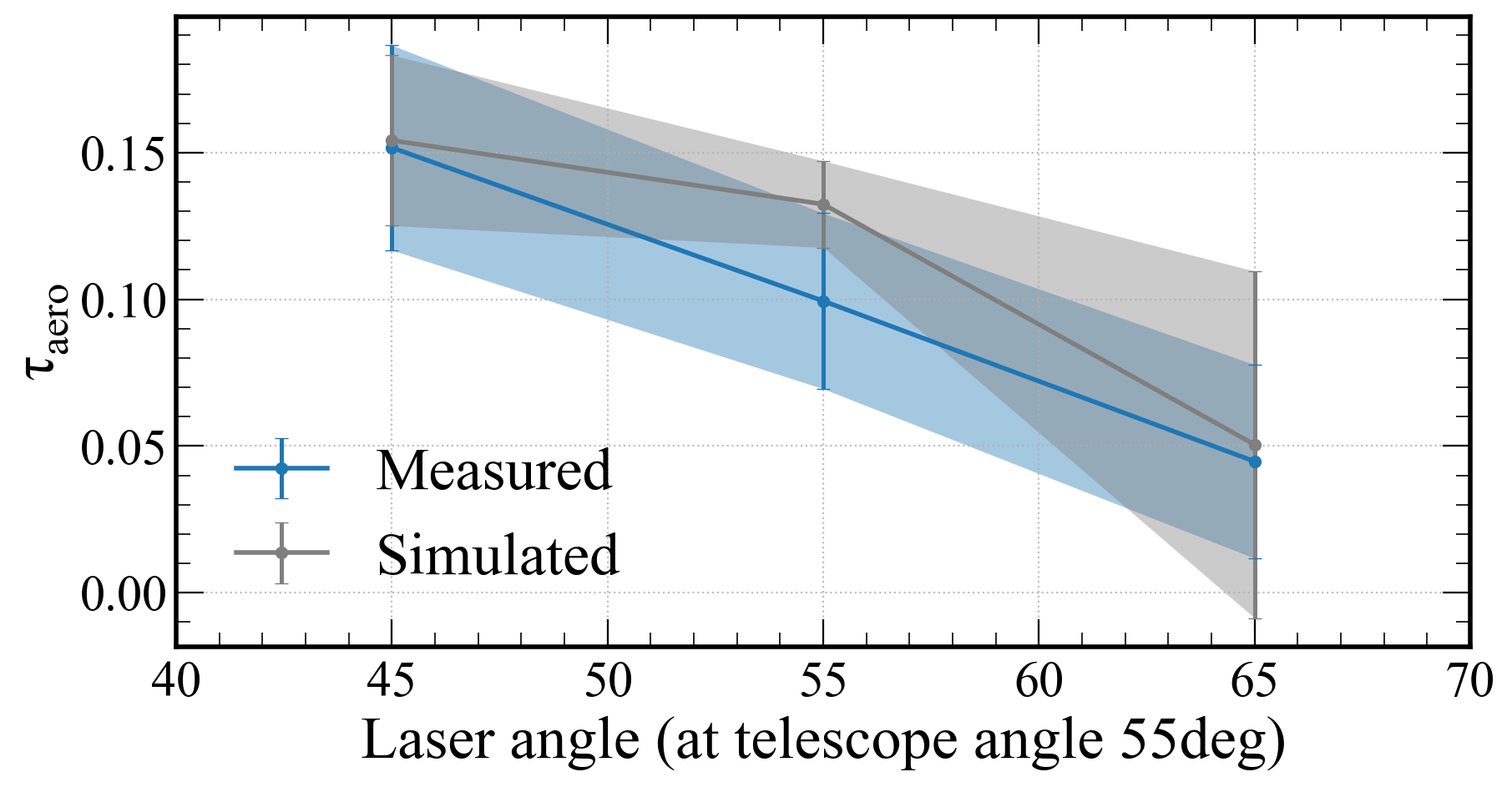}
\caption{Laser track with a laser elevation angle of 55\degree towards the telescope pointing at 55\degree. The section of the laser track is above the aerosol layer. This allows for the determination of vertical aerosol optical depth.}\label{fig:aero}
\end{figure}

More measurements with the laser included constant laser energy and azimuth pointing directly toward each other, where the laser aimed at the center of the camera. By changing the elevation angle of the laser and the telescope, this setup is a bi-dynamical Lidar system, with the FT acting as an orientable receiver. Measurements of the laser in the laser-detector plane or shower-detector-plane (SDP) are used to determine the angular resolution of the FT, see Fig. \ref{fig:angle}.
And a novel diagnostic was explored to measure the vertical optical depth of the aerosol in the atmosphere, see Fig. \ref{fig:aero}. The advantage of the new aerosol study is that an absolute calibration of the telescope is not necessary. \\

Additionally, there were measurements of the night sky background for a successfully performed flat fielding and star measurements that allow for absolute calibration of the FT using star catalogs. Other measurements included meteors, LEDs, and comic rays.

\newpage
\subsection{Conclusion}
The EUSO-SPB2 fluorescence telescope calibration and field tests were successfully conducted, showcasing an enhancement in optical efficiency and focus compared to EUSO-SPB1, nearly doubling its performance. The application of a high-power UV laser proved to be a reliable and sustainable technique for optical telescope calibration and astroparticle experiments. Subsequently, the EUSO-SPB2 telescope was launched successfully and functioned as intended, efficiently measuring data during its airborne operation. Despite the limited flight opportunity, this achievement represents a significant milestone toward space-based Ultra-High-Energy Cosmic Ray (UHECR) and Very-High-Energy (VHE) neutrino observatories.

\subsection{Acknowledgement}
The authors acknowledge the support by NASA awards 11-APRA-0058, 16-APROBES16-0023, 17-APRA17-0066, NNX17AJ82G, NNX13AH54G, 80NSSC18K0246, 80NSSC18K0473, 80NSSC19K0626, 80NSSC18K0464, 80NSSC22K1488, 80NSSC19K0627 and 80NSSC22K0426, 
the French space agency CNES, 
National Science Centre in Poland grant n. 2017/27/B/ST9/02162, 
and by ASI-INFN agreement n. 2021-8-HH.0 and its amendments. 
This research used resources of the US National Energy Research Scientific Computing Center (NERSC), the DOE Science User Facility operated under Contract No. DE-AC02-05CH11231. 
We acknowledge the NASA BPO and CSBF staff for their extensive support. 
We also acknowledge the invaluable contributions of the administrative and technical staff at our home institutions.

\bibliography{my-bib-database}

\providecommand{\href}[2]{#2}\begingroup\raggedright\begin{thebibliography}{1}

\bibitem{SPB1}
J.~Eser, \emph{Results of the euso-spb1 flight},  in \emph{36th International
  Cosmic Ray Conference}, vol.~358, p.~247, 07, 2019,
  \href{https://doi.org/10.22323/1.358.0247}{DOI}.

\bibitem{Olinto_2021}
P.~collaboration, A.~Olinto, J.~Krizmanic, J.~Adams, R.~Aloisio et~al.,
  \emph{The poemma (probe of extreme multi-messenger astrophysics)
  observatory},
  \href{https://doi.org/10.1088/1475-7516/2021/06/007}{\emph{Journal of
  Cosmology and Astroparticle Physics} {\bfseries 2021} (2021) 007}.

\bibitem{MAPMT}
Hamamatsu, ``Photomultiplier tubes (pmts).''
  \url{https://www.hamamatsu.com/us/en/product/optical-sensors/pmt.html}, 2023.

\end{thebibliography}\endgroup

\newpage
{\Large\bf Full Authors list: The JEM-EUSO Collaboration\\}
%{\scriptsize (author-list as of July 15th, 2023 with reorganized affiliations)} \hspace{0.6cm}
%{\scriptsize (version  \today{} \currenttime{})}
%\vspace*{0.5cm}

\begin{sloppypar}
{\small \noindent
S.~Abe$^{ff}$, 
J.H.~Adams Jr.$^{ld}$, 
D.~Allard$^{cb}$,
P.~Alldredge$^{ld}$,
R.~Aloisio$^{ep}$,
L.~Anchordoqui$^{le}$,
A.~Anzalone$^{ed,eh}$, 
E.~Arnone$^{ek,el}$,
M.~Bagheri$^{lh}$,
B.~Baret$^{cb}$
% ...
}
\end{sloppypar}
\vspace*{.3cm}

%%\newpage
{ \footnotesize
\noindent
% Algeria - 2 institutes
% ...
$^{cb}$ Universit\'e de Paris, CNRS, AstroParticule et Cosmologie, F-75013 Paris, France\\
% ...
$^{ed}$ Istituto Nazionale di Fisica Nucleare - Sezione di Catania, Italy\\
% ...
$^{eh}$ INAF - Istituto di Astrofisica Spaziale e Fisica Cosmica di Palermo, Italy\\
% ...
$^{ek}$ Istituto Nazionale di Fisica Nucleare - Sezione di Torino, Italy\\
$^{el}$ Dipartimento di Fisica, Universit\`a di Torino, Italy\\
% ...
$^{ep}$ Gran Sasso Science Institute, L'Aquila, Italy\\
% Japan - 7 institutes 
% ...
$^{ff}$ Nihon University Chiyoda, Tokyo, Japan\\ 
% ...
$^{ld}$ University of Alabama in Huntsville, Huntsville, AL, USA\\
$^{le}$ Lehman College, City University of New York (CUNY), NY, USA\\
% ...
$^{lh}$ Georgia Institute of Technology, USA\\
% ...
}

%\begin{thebibliography}{99}
%\bibitem{...}
%....

%\end{thebibliography}

%% Full authors list (ONLY FOR COLLABORATIONS)
%\clearpage
%\section*{Full Authors List: \Coll\ Collaboration}
%
%\noindent \textbf{Note comment afterwards:} Collaborations have the possibility to provide an authors list in xml format which will be used while generating the DOI entries making the full authors list searchable in databases like Inspire HEP. \\
%
%\scriptsize
%\noindent
%first.author$^1$, 
%second.author$^2$, 
%third.author$^3$ % .... more names
%and 
%last.author$^{n}$ \\
%
%\noindent
%$^1$first.affiliation.
%$^2$second.affiliation. % .... more affiliation
%$^{m}$last.affiliation.

\end{document}